\documentclass[draft]{agujournal2018}
\draftfalse
\usepackage{url}
\usepackage{lineno}
\usepackage{booktabs}
\usepackage{amssymb}

\usepackage{journal_macros}

\usepackage[bookmarks=false, pdfnewwindow=true, colorlinks=true, linkcolor=magenta, citecolor=blue, filecolor=cyan, urlcolor=purple, final=true]{hyperref}

\journalname{Space Weather}

\newcommand\editone[1]{\textcolor{black}{#1}}
\newcommand\edittwo[1]{\textcolor{black}{#1}}

\graphicspath{{./}{figures/}}

\begin{document}

\title{Monitoring the Solar Wind Before It Reaches L1}

\authors{Erika~Palmerio\affil{1}}

\affiliation{1}{Predictive Science Inc., San Diego, CA, USA}

\correspondingauthor{E.~Palmerio}{epalmerio@predsci.com}

%%%%%%%%%%%%%%%%%%%%%%%%%%%%%%%%%%%%%%%%
%%% KEYPOINTS ** 140 characters max each **
%%%%%%%%%%%%%%%%%%%%%%%%%%%%%%%%%%%%%%%%

\begin{keypoints}
\item Space weather predictions of the solar wind usually rely on in-situ measurements at the L1 point, with lead times of a few tens of minutes
\item The G5 storm of 2024 May provided an opportunity to test real-time predictions using STEREO-A data, 0.04~au closer to the Sun than L1
\item We discuss the recent results by Weiler et al.\ (2025) in the context of monitoring the solar wind upstream of L1 for improved predictions
\end{keypoints}

%%%%%%%%%%%%%%%%%%%%%%%%%%%%%%%%%%%%%%%%
%%% ABSTRACT ** Max 250 words **
%%%%%%%%%%%%%%%%%%%%%%%%%%%%%%%%%%%%%%%%

\begin{abstract}
Space weather predictions of the solar wind impacting Earth are usually first based on remote-sensing observations of the solar disc and corona, and eventually validated and/or refined with in-situ measurements taken at the Sun--Earth Lagrange L1 point, where real-time monitoring probes are located. However, this pipeline provides, on average, only a few tens of minutes of lead time, which decreases to ${\sim}30$~minutes or less for large solar wind speeds of ${\sim}800$~km$\cdot$s$^{-1}$ and above. The G5 geomagnetic storm of 2024 May provided an opportunity to test predictions generated employing real-time data from the STEREO-A spacecraft, placed $13^{\circ}$ west of Earth and 0.04~au closer to the Sun than L1 at the time of the event, as shown recently by \citet{weiler2025}. In this Commentary, we contextualise these results to reflect upon the advantages of measuring the solar wind in situ upstream of L1, leading to improvements in both fundamental research of interplanetary physics and space weather predictions of the near-Earth environment.
\end{abstract}

%%%%%%%%%%%%%%%%%%%%%%%%%%%%%%%%%%%%%%%%
%%% PLAIN LANGUAGE SUMMARY ** Max 200 words **
%%%%%%%%%%%%%%%%%%%%%%%%%%%%%%%%%%%%%%%%

\begin{plainlanguagesummary}
The solar wind, streaming continuously outwards from the Sun, can embed structures that have the potential to cause significant space weather effects at and near Earth, such as coronal mass ejections. Currently, the solar wind impacting Earth is estimated with a relatively high degree of confidence only after its properties are directly measured at the Lagrange L1 point, located 0.01~au ahead of Earth towards the Sun and where a handful of satellites monitor the local conditions. This results in space weather forecast leading times of the order of a few tens of minutes. \citet{weiler2025} took advantage of a fortuitous spacecraft configuration in 2024 May and the strongest geomagnetic storm in two decades to demonstrate how warning times can be shortened by employing real-time measurements from probes placed closer to the Sun than L1. In this Commentary, we summarise these results and discuss advantages and challenges of measuring the solar wind from so-called ``sub-L1 monitors''.
\end{plainlanguagesummary}

%%%%%%%%%%%%%%%%%%%%%%%%%%%%%%%%%%%%%%%%
%%% INTRODUCTION
%%%%%%%%%%%%%%%%%%%%%%%%%%%%%%%%%%%%%%%%

\section{Introduction} \label{sec:intro}

The near-Earth space environment is constantly shaped and affected by the dynamic solar activity, resulting in a plethora of phenomena both on ground and in space collectively known as space weather. Amongst the solar drivers that can induce disturbances in geospace is the ever-blowing solar wind, which has been monitored continuously from the Sun--Earth Lagrange L1 point (0.01~au ahead of Earth from the Sun) since the 1990s---initially with science missions such as the Solar and Heliospheric Observatory \citep[SOHO;][]{domingo1995}, the Advanced Composition Explorer \citep[ACE;][]{stone1998}, and Wind \citep{ogilvie1997}, which were later joined by the first deep-space operational spacecraft, i.e.\ the Deep Space Climate Observatory \citep[DSCOVR;][]{burt2012}. The L1 point is a crucial location for space weather monitoring, allowing probes to measure the solar wind locally before it impacts Earth's magnetosphere whilst maintaining a relatively low fuel consumption. The importance of L1 for operational forecasting and mitigation is remarked further by the fact that there are plans in place to maintain assets at this location for the foreseeable future, with the Space Weather Follow On L1 (SWFO-L1) \edittwo{recently launched in} \editone{September} 2025.

Nevertheless, measuring the solar wind at L1 provides only a brief warning time before a given structure eventually impacts Earth, ranging from just under 1.5 hours for slower flows of ${\sim}300$~km$\cdot$s$^{-1}$ to approximately 25 minutes for more extreme speeds of ${\sim}1000$~km$\cdot$s$^{-1}$. These lead times are much shorter than the ideal time ranges of a few hours to a couple of days required by end-users \citep[e.g.,][]{jackson2023, vourlidas2023}. One approach that allows to provide longer-term forecasts is via employing global, magnetohydrodynamics (MHD) modelling of the ``background'' solar wind and its transient structures such as coronal mass ejections (CMEs), as has been done operationally for over a decade \citep[e.g.,][]{pizzo2011}. However, uncertainties in modelling the ambient medium \citep[e.g.,][]{gressl2014, reiss2023} as well as in determining CME input parameters \citep[e.g.,][]{kay2024a, verbeke2023} translate directly into advance forecasts that are characterised by relatively large uncertainties \citep[see, e.g.,][]{jian2015, kay2024b}. An additional challenge is provided by predicting the north--south magnetic field component ($B_{Z}$) embedded in CMEs \citep[e.g.,][]{kilpua2019, vourlidas2019}, which is strongly correlated to the corresponding geomagnetic response and that is currently not forecast operationally before a given structure is measured directly at L1. These issues are amongst the reasons that have pushed the research community towards reflecting upon the utility of measuring the solar wind in situ before it reaches L1, i.e.\ via sub-L1 monitors \citep[e.g.,][]{lugaz2025, morley2020, nasem2025}.

The recent 2024 May G5 geomagnetic storm \citep[e.g.,][]{hayakawa2025} provided an opportunity to test in a real-time-like scenario how sub-L1 measurements improve the warning time of solar wind forecasts. \citet{weiler2025} employed data available in real time from the Solar Terrestrial Relations Observatory Ahead \citep[STEREO-A;][]{kaiser2008}, located 0.044~au \editone{upstream} and $12.5^{\circ}$ west of L1 at the time of the event, to \editone{show that such a sub-L1 monitor would have allowed extension of the} prediction lead time by approximately 2.5~hours. In this Commentary, we first provide a brief summary of the approach and results reported by \citet{weiler2025}, and then discuss the advantages and challenges of monitoring the solar wind upstream of L1 in the context of space weather research as well as operations.

%%%%%%%%%%%%%%%%%%%%%%%%%%%%%%%%%%%%%%%%
%%% SUMMARY OF WEILER ET AL. (2025)
%%%%%%%%%%%%%%%%%%%%%%%%%%%%%%%%%%%%%%%%

\section{Catching a Storm Before It Impacts L1: The 2024 May Event} \label{sec:may2024}

\begin{figure}[th!]
\centering
\includegraphics[width=\linewidth]{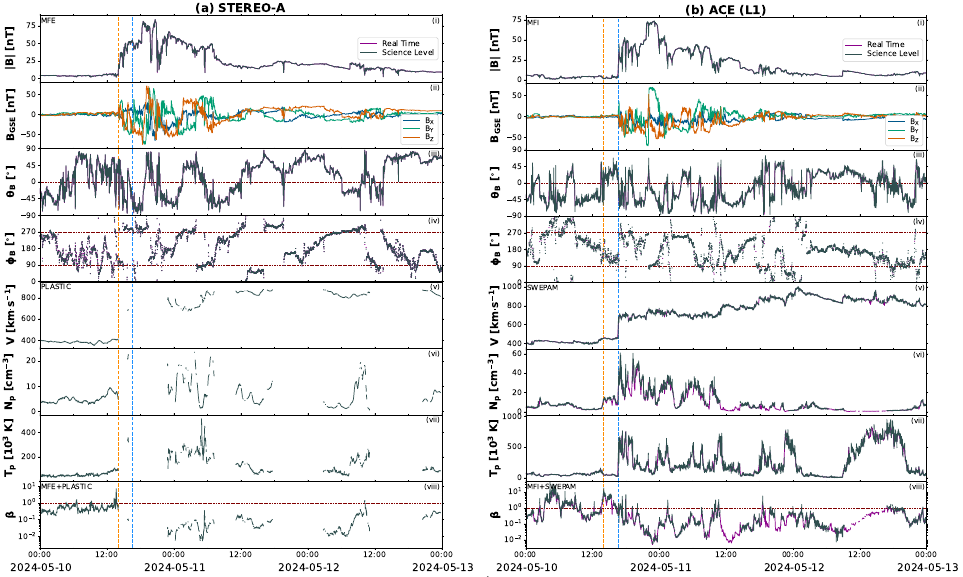}
\caption{Overview of the available in-situ observations near 1~au of the interplanetary structure(s) responsible for the 2024 May geomagnetic storm, showing data collected by (a) STEREO-A and (b) ACE. Each panel shows, from top to bottom: (i) magnetic field magnitude, (ii) Cartesian fields components in Geocentric Solar Ecliptic (GSE) coordinates, (iii) latitudinal and (iv) longitudinal angles of the magnetic field, (v) solar wind speed, (vi) proton density, (vii) proton temperature, and (viii) plasma beta. Note that plasma data from STEREO-A were not available in real time, whilst the remaining data sets (magnetic field from STEREO-A as well as magnetic field and plasma from ACE) \editone{are displayed in both (purple) real-time and (grey) science-level formats (except for panel (ii), where only science-level data are shown)}. In both panels, the orange and \editone{blue} dashed lines mark the first shock arrival time at STEREO-A and ACE, respectively.}
\label{fig:may2024}
\end{figure}

The 2024 May ``superstorm''---\editone{mentioned in the literature also as the Mother's Day storm \citep[e.g.,][]{kruparova2024} and/or the Gannon storm \citep[e.g.,][]{thampi2025}, in honour of Jennifer Gannon \citep{lugaz2024b}}---was the strongest since the 2003 ``Halloween'' event, reaching a Kp index of 9 (resulting in a storm of G5 category) and a Dst index of $-406$~nT. It gained significant attention in the research community \citep[e.g.,][]{hajra2024, hayakawa2025, jaswal2025, liu2024}, and the overall consensus is that the observed geomagnetic effects were caused by at least five CMEs launched in close succession and interacting in interplanetary space. An interesting feature of this event was that the STEREO-A spacecraft was located upstream of and relatively close to the L1 point, i.e.\ 0.044~au \editone{upstream} and $12.5^{\circ}$ to the west. This fortuitous positioning was exploited by \citet{weiler2025} to formulate predictions of the interplanetary magnetic fields impacting Earth and geomagnetic indices in a real-time-like scenario, i.e.\ employing uniquely information and data available at the time of the event. An overview of the in-situ observations near L1 from the STEREO-A and Advanced Composition Explorer \citep[ACE;][]{stone1998} spacecraft is provided in Figure~\ref{fig:may2024}.

\citet{weiler2025} took advantage of the earlier (by ${\sim}2.5$~hours, as highlighted in Figure~\ref{fig:may2024}) arrival of the first disturbance---i.e., an interplanetary shock---at STEREO-A to test and demonstrate an extension of the prediction lead time. Apart from modelling the arrival times and speeds of the five CMEs under consideration based on remote-sensing observations using the ELliptical Evolution \citep[ELEvo;][]{mostl2015} model, the authors employed real-time magnetic field data from STEREO-A (no plasma measurements were available at the time of the event) to predict the geomagnetic effects of the complex series of interacting transients. By applying the empirical formulas by \citet{temerin2006}, \editone{which employ solar wind data to assess the corresponding geomagnetic response in terms of the Dst index}, and combining STEREO-A magnetic field measurements with estimates of the plasma parameters, \citet{weiler2025} predicted Dst and SYM‐H minimum values of $-460 \pm 55$~nT and $-479 \pm 8$~nT, respectively, overestimating the observed minimum Dst index of $-406$~nT by 12\% and underestimating the observed SYM‐H of $-518$ nT by 8\%. Considering that the predictions could be generated starting ${\sim}2.5$~hours before the first impact at L1, these results appear rather encouraging and provide \editone{an illustration of the potential utility of a longer lead time, as well as} a first assessment of the feasibility of sub-L1 monitoring in the context of a ``worst-case scenario'' (i.e.\ a G5-class storm driven by multiple, interacting CMEs).

Additionally, it is intuitive to assume that the hindcasts reported by \citet{weiler2025} may have proven even more accurate had STEREO-A plasma data been available in real time. The plasma parameters that were released after the event (displayed in Figure~\ref{fig:may2024}), despite being characterised by significant data gaps, show a solar wind speed spanning a range of ${\sim}200$~km$\cdot$s$^{-1}$ and a highly variable density profile, suggesting that the constant, average values estimated by \citet{weiler2025} may not have accurately described the finer structure of the disturbances encountered. On the other hand, concluding that better measurements at STEREO-A should have \editone{yielded} better predictions of the solar wind later impacting ACE may be too simplistic of an assumption, which neglects CME evolution with radial distance and structural differences over a given angular separation. In fact, in the case of the 2024 May storm, \citet{liu2024} and \citet{weiler2025} agreed that the \editone{estimated} geomagnetic impact at \editone{STEREO‐A's heliolongitude would} have been higher than at L1, given that the CMEs were more directed towards STEREO‐A. \editone{In this context, the extent to which real-time plasma data improve predictions---and, more critically, whether this improvement depends on the angular separation between the sub-L1 monitor and Earth---is yet to be quantified.} Having established that an operational solar wind monitor should be equipped at the very least with magnetic field and plasma instruments \editone{designed to be resilient during extreme events}, questions remain as to how many sub-L1 probes are necessary for reliable predictions and their optimal location(s) upstream of Earth. In the next section, we explore these issues in light of existing studies and proposed mission concepts.

%%%%%%%%%%%%%%%%%%%%%%%%%%%%%%%%%%%%%%%%
%%% MONITORS UPSTREAM OF L1
%%%%%%%%%%%%%%%%%%%%%%%%%%%%%%%%%%%%%%%%

\section{Advantages and Challenges of Solar Wind Monitoring Upstream of L1} \label{sec:upstream}

\begin{figure}[th!]
\centering
\includegraphics[width=0.65\linewidth]{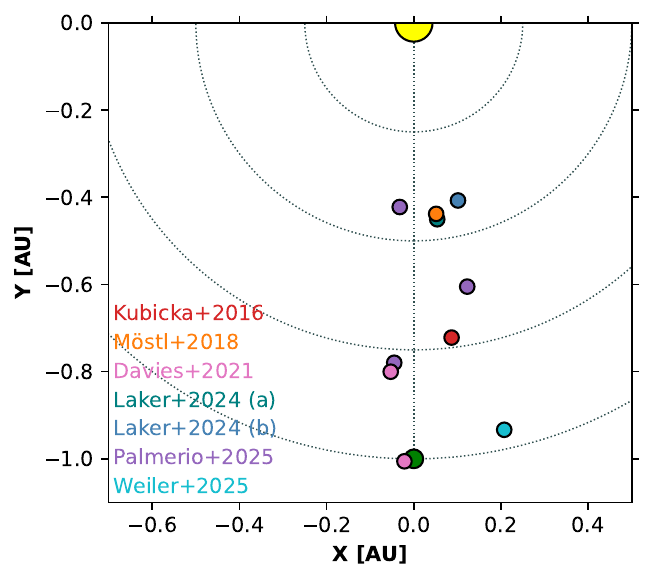}
\caption{Visualisation illustrating the upstream monitor positions that characterised the ``sub-L1 CME hindcasting'' studies discussed in this Commentary. The positions are reported at the time of the corresponding CME shock arrival at 1~au, and each colour represents a different event \citep[note that the study of][considers two separate cases]{laker2024}. The larger green circle represents the location near 1~au considered in each work \citep[i.e., STEREO-A in the analysis of][and L1 in the remaining studies]{palmerio2025}.}
\label{fig:orbits}
\end{figure}

Given that the purpose of a sub-L1 monitor is, by design, to measure the solar wind somewhere below 0.99~au to provide earlier warnings than from L1, one of the most critical aspects to consider is where such a probe should be placed in terms of heliocentric distance. In fact, a spacecraft to be employed in real-time operations should achieve ``the perfect balance'' between being as close as possible to the Sun (to detect incoming structures as soon as possible) and being as close as possible to Earth (to ensure that the structures encountered have not evolved dramatically by the time they reach 1~au). Some insights on the matter are provided by previous studies that, in a hindcast fashion, have taken advantage of fortuitous radial alignments of an asset at 1~au with an inner spacecraft (see Figure~\ref{fig:orbits}). 
\edittwo{For example, \citet{kubicka2016} used Venus data to forecast a CME--CME interaction event, achieving satisfactory agreement in arrival time (${\sim}6$~hours late), speed (${\sim}530$ vs.\ ${\sim}490$~km$\cdot$s$^{-1}$), and Dst ($-100$ vs.\ $-71$~nT) with a 21-hour prediction lead time. \citet{mostl2018} leveraged in-situ data at Mercury to constrain CME flux rope modelling up to Earth, slightly underestimating both speed (by ${\sim}50$~km$\cdot$s$^{-1}$) and Dst index ($-50$ vs.\ $-73$~nT). \citet{davies2021} employed data at ${\sim}0.8$~au as an upstream monitor, reproducing the Dst index ($-47$ vs.\ $-60$~nT) despite find a slower-than-expected decay in magnetic field strength. \citet{laker2024} showed that measurements near $0.5$~au reduced CME arrival time errors by factors of two to four across two events, noting that one case preserved a similar magnetic configuration at Earth whereas the other was altered by CME--CME interaction. Finally, \citet{palmerio2025} analysed a CME encountered by four spacecraft (at 0.4, 0.6, 0.8, and 1~au) to assess the use of inner measurements for predictions, finding consistent results between 0.8 and 1~au but larger discrepancies at smaller heliocentric distances. Although limited in number and radial coverage, these studies collectively} suggest that any spacecraft inside 1~au close to radial alignment with Earth should provide \editone{some degree of} improvement in the accuracy of space weather forecasts.

Another challenge is represented by whether a single monitor aligned with the Sun--Earth line is sufficient for reliable forecasts, or whether multiple probes should be placed at given angular separation(s) to provide a more robust range of uncertainties for predictions\editone{---an aspect that presently can only be addressed through probabilistic modelling \citep{morley2018, obrien2023}}. In fact, \citet{borovsky2018} noted that even measurements taken at L1 are not fully representative of the solar wind impacting Earth's magnetosphere, as e.g.\ the clock angle of the interplanetary magnetic field is highly dependent on the traversing structure. Studies that took advantage of small spacecraft separations in near-Earth space reported that it is possible to find non-negligible differences between two data sets even for distances below ${\sim}0.01$~au \citep{ala-lahti2020, koval2010, lugaz2018}. Even more so, dramatic variations between two sets of measurements characterised by an angular separation ${<}5^{\circ}$ were reported even for a CME encountered at Mercury's orbit, i.e.\ at approximately one-third of the Sun--Earth distance, where structures in the solar wind are expected to have undergone less evolutionary processes \citep{palmerio2024}. These works suggest that a small constellation of sub-L1 monitors may provide a more complete assessment of the solar wind that is expected to later impact L1 and Earth's magnetosphere, \editone{including uncertainties \citep[see also the discussion in][]{lugaz2025}}.

Of course, establishing the optimal location(s) of sub-L1 monitor(s) is only one of the hurdles to overcome, another important aspect to consider being the feasibility of placing one or more space weather sentinels somewhere inside 1~au in terms of mission design. Several concepts have been proposed over the past couple of decades, ranging from four spacecraft in distance retrograde orbits at 0.9~au \citep{stcyr2000}, to six spacecraft in a planet-like orbit at 0.72~au \citep{ritter2015}, to multiple probes sampling the solar wind below L1 along the Sun--Earth line \citep{akhavan-tafti2023}, to a small constellation orbiting Earth that reaches up to ${\sim}0.04$~au upstream of L1 \editone{\citep{lugaz2024c}}, amongst others. What all these proposed missions have in common is that they involve multiple probes, thus highlighting that spacecraft constellations are not only desirable from a research and operational perspective, but would also ensure that at least one spacecraft is always ``close enough'' to the Sun--Earth line due to orbital constraints. Ultimately, a sub-L1 monitoring mission will have to achieve an optimal combination of number of probes involved, heliocentric distance(s) covered, and maximum angular separation to the Sun--Earth line, ideally motivated by insights from research efforts such as the works summarised above as well as funding availability.

%%%%%%%%%%%%%%%%%%%%%%%%%%%%%%%%%%%%%%%%
%%% CONCLUDING REMARKS
%%%%%%%%%%%%%%%%%%%%%%%%%%%%%%%%%%%%%%%%

\section{Concluding Remarks} \label{sec:conclusion}

In this Commentary, we have reflected upon the importance of employing sub-L1 monitors for improved real-time space weather forecasts. Motivated by the recent results of \citet{weiler2025}, who evaluated how taking advantage of measurements upstream of L1 improves the prediction lead time in the context of the 2024 May G5 storm, we have provided a brief overview of the advantages and challenges of designing an \editone{operational} mission aimed at sampling the solar wind before it reaches L1. \editone{Existing works showed} that any monitor close to radial alignment with Earth at sub-L1 distances should provide at least some improvement to forecasts and that multi-spacecraft missions are likely to supply more reliable predictions, but a larger statistics is needed to determine the optimal positioning (in terms of heliocentric distance(s) and angular separation(s) with L1 and Earth). For example, many multi-event studies have investigated how solar wind transients evolve with radial distance \citep[e.g.,][]{good2019, salman2020, vrsnak2019} and vary across more or less large longitudinal separations \citep[e.g.,][]{banu2025, lugaz2024a, lugaz2025}, but only a handful of them employed sub-L1 measurements in a forecasting-like scenario (see Section~\ref{sec:upstream} and Figure~\ref{fig:orbits}). Revisiting these existing data sets ``as if they were available in real time'' may shed additional light in this regard. It is possible that there may be an optimal distance/separation for improved arrival time estimates and a different one for magnetic field (and $B_{Z}$) predictions, in which case a balance between the different parameters to forecast together with mission design attainability will have to be reached.

Nonetheless, it is clear that the space weather community is progressively converging towards a consensus about the necessity and feasibility of upstream real-time monitors, and is beginning to explore potential implementation avenues and opportunities. There is no planned sub-L1 endeavour at the time of writing, but should such a mission occur, there is promise not only for significant improvements in space weather forecasting \editone{(including more accurate predictions of both the incoming solar wind and its impacts on the geospace environment)}, but also for deepening our knowledge of how the solar wind and its transient events evolve and/or vary across different temporal and spatial scales.

%%%%%%%%%%%%%%%%%%%%%%%%%%%%%%%%%%%%%%%%
%%% CONCLUDING REMARKS
%%%%%%%%%%%%%%%%%%%%%%%%%%%%%%%%%%%%%%%%

\section*{Data Availability Statement}

The definitive Kp index is provided by \edittwo{GFZ Potsdam (accessible at \url{https://doi.org/10.5880/Kp.0001})}.
The provisional Dst index is provided by the World Data Center for Geomagnetism, Kyoto (accessible at \url{https://wdc.kugi.kyoto-u.ac.jp}).
STEREO-A and ACE in-situ data are archived at NASA's Coordinated Data Analysis Web (CDAWeb; accessible at \url{https://cdaweb.gsfc.nasa.gov}).

%%%%%%%%%%%%%%%%%%%%%%%%%%%%%%%%%%%%%%%%
%%% ACKNOWLEDGEMENTS
%%%%%%%%%%%%%%%%%%%%%%%%%%%%%%%%%%%%%%%%

\acknowledgments
E.~Palmerio acknowledges support from NASA's Heliophysics Guest Investigators-Open (grant no.\ 80NSSC23K0447) and Living With a Star (grant no.\ 80NSSC24K1108) programmes.

\bibliography{bibliography}

\end{document}